\documentclass{article}

\usepackage[dblblindworkshop, final]{neurips_ACA_2025}

\workshoptitle{\\ Workshop on Algorithmic Collective Action (ACA@NeurIPS 2025)}

\usepackage[utf8]{inputenc} 
\usepackage[T1]{fontenc}    
\usepackage{hyperref}       
\usepackage{url}            
\usepackage{booktabs}       
\usepackage{amsfonts}       
\usepackage{nicefrac}       
\usepackage{microtype}      
\usepackage{xcolor}         

\usepackage{amsmath}
\usepackage{amsfonts}
\usepackage{graphicx}

\usepackage{algorithm}
\usepackage{algorithmicx}
\usepackage{algpseudocode}

\title{Measuring the Hidden Cost of Data Valuation through Collective Disclosure}

\author{%
  Patrick Mesana \\
  HEC Montréal \\
  Montréal, Québec, Canada \\
  \texttt{patrick.mesana@hec.ca} \\
  \And
  Gilles Caporossi \\
  HEC Montréal \\
  Montréal, Québec, Canada \\
  \texttt{gilles.caporossi@hec.ca} \\
  \And
  Sébastien Gambs \\
  Université du Québec à Montréal \\
  Montréal, Québec, Canada \\
  \texttt{sebastien.gambs@uqam.ca} \\
}

\begin{document}

\maketitle

\begin{abstract}
Data valuation methods assign marginal utility to each data point that has contributed to the training of a machine learning model. 
If used directly as a payout mechanism, this creates a \emph{hidden cost of valuation}, in which contributors with near-zero marginal value would receive nothing, even though their data had to be collected and assessed. 
To better formalize this cost, we introduce a conceptual and game-theoretic model—the \textit{Information Disclosure Game}—between a \textit{Data Union} (sometimes also called a data trust), a member-run agent representing contributors, and a \textit{Data Consumer} (\emph{e.g.}, a platform).
After first aggregating members’ data, the DU releases information progressively by adding Laplacian noise under a differentially-private mechanism. 
Through simulations with strategies guided by data Shapley values and multi-armed bandit exploration, we demonstrate on a Yelp review helpfulness prediction task that data valuation inherently incurs an explicit acquisition cost and that the DU’s collective disclosure policy changes how this cost is distributed across members.
\end{abstract}

\section{Introduction}

The idea of data dividends~\citep{feygin2021data}—distributing a share of the value generated from data back to contributors—has drawn attention as a mechanism to make data economies more inclusive. 
A way to ground such dividends is through data valuation, in which each point’s contribution to predictive performance is quantified. 
Recent methods such as data Shapley value (DSV)~\citep{jia2019towards}, which originated from cooperative game theory, have been applied to data summarization and efficient acquisition~\citep{ghorbani2019data}. 
These methods show that only a fraction of the data is needed for high utility, but nonetheless require access to the entire dataset to compute the marginal contributions—meaning that each individual must first contribute their data, even if their eventual payout is negligible. 
We coin this as the \emph{hidden cost of data valuation}.

Following calls for data trusts and unions as institutional forms of collective governance~\citep{delacroix2019bottom}, we frame our approach as a form of algorithmic collective action in which contributors act collectively through a Data Union (DU) that sets group data disclosure policies, shifting how value and costs are distributed. 
While recent work shows that differentially-private mechanisms ~\cite{dwork2006differential} can diminish the leverage of collective action in gradient-based training~\citep{solanki2025crowding}, we instead use DP positively as a tool for disclosure and value control.

\begin{figure*}[h!]
    \centering
    \includegraphics[width=1\textwidth]{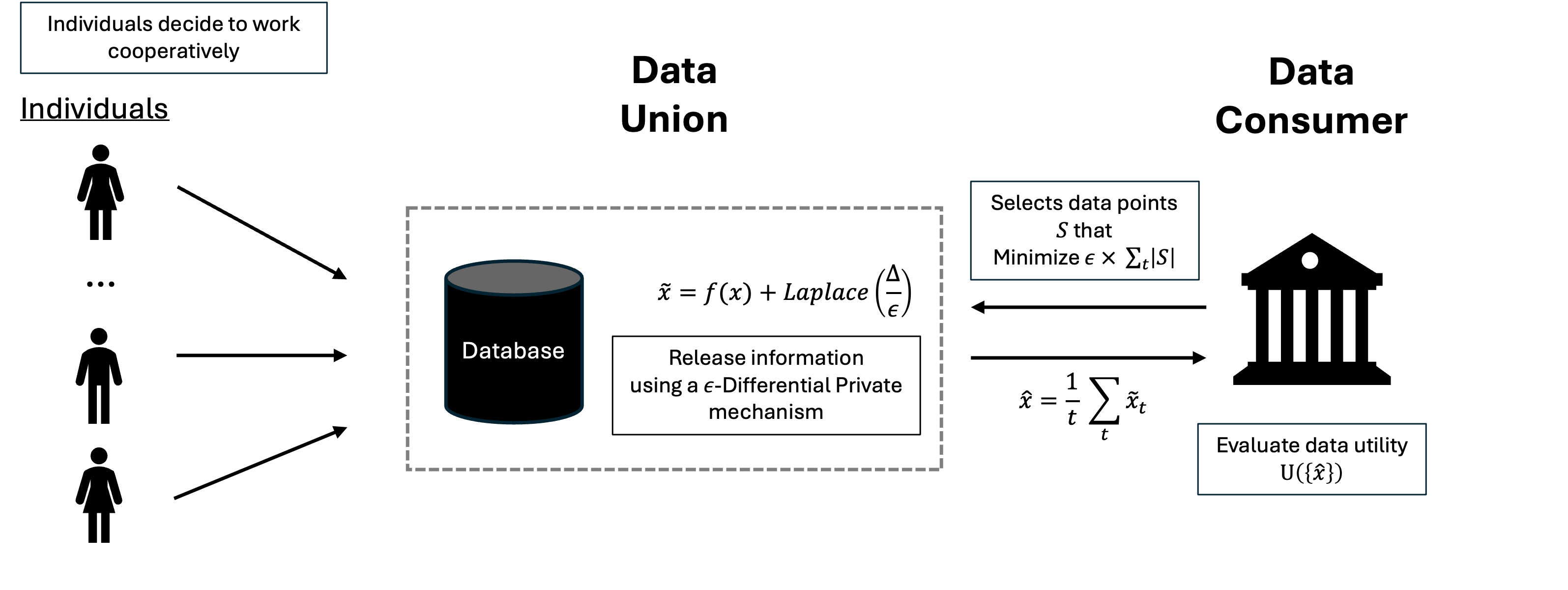}
    \caption{Illustration of the Information Disclosure Game. 
    A Data Union (DU) holds a private dataset and releases information to a Data Consumer (DC) by adding Laplacian noise to data points under $\epsilon$-differential privacy. 
    The DC incrementally acquires these noisy version of the data points, denoises them using an average and train a model to reach a utility target. In this work, we focus on non-parametric $k$-Nearest Neighbors ($k$-NN) and SBERT embeddings \cite{reimers-2019-sentence-bert}.}
    \label{fig:info_disclosure_game}
\end{figure*}

More precisely, we introduce a conceptual model in which contributors coordinate through a DU.
Rather than acting individually in a decentralized marketplace, members pool their data and empower the union to negotiate collectively. 
The DU aggregates the dataset and controls how information is released to a Data Consumer (DC) by progressively disclosing noisy versions of data points using a differentially-private mechanism. 
In this way, the DU operates as a collective agent: it sets disclosure policies that influence which points the DC must acquire to reach its utility target, and thus how dividends are distributed across members. 
See Figure~\ref{fig:info_disclosure_game} for an illustration of our approach.

We formalize this interaction as an Information Disclosure Game (IDG), a form of  Stackelberg game~\cite{simaan1973stackelberg, fudenberg1991game} in which the DU acts as the leader by choosing disclosure rules, and the DC acts as the follower by optimizing acquisitions given those rules (Figure~\ref{fig:info_disclosure_game}). 
This framework allows for the explicit identification of the acquisition cost that valuation imposes on the DC and how collective disclosure policies shape their distribution across members.
We instantiate the model through a $k$-Nearest Neighbors ($k$-NN), in which the relationship between individual points and predictive utility is direct. 
Simulating the DC strategies guided by data Shapley values and by multi-armed bandit exploration, we evaluated it on a Yelp review helpfulness prediction task using text embeddings.

\paragraph{Outline.} First in Section~\ref{sec:background}, we review the related work on data valuation and differential privacy before detailing in Section~\ref{sec:framework} the IDG framework and the objectives of DU and DC. Afterwards, in Section~\ref{sec:case_study}, we present our empirical evaluation followed by a discussion and conclusion.

\section{Background}
\label{sec:background}

\paragraph{Data valuation.}
In cooperative game theory, an \emph{imputation} is a way to distribute the total value of a game among its players~\cite{osborne1994course}. 
In the context of data valuation, each data point in a dataset \(D\) is treated as a \emph{player} and a \emph{value function} \(v : 2^D \to \mathbb{R}\) is used to the utility (\emph{e.g.}, accuracy or loss reduction) of any subset \(D' \subseteq D\). 
The objective is to find an imputation that allocates fairly the total value \(v(D)\) among individual data points based on their contributions. 
Several approaches have been explored for data valuation---each with different fairness, computational and interpretability trade-offs---including \emph{Data Shapley}~\cite{ghorbani2019data, jia2019towards}, \emph{Data Banzhaf}~\cite{wangDataBanzhafRobust2023}, \emph{Beta Shapley}~\cite{kwonBetaShapleyUnified2022} and the \emph{Core}~\cite{yanIfYouShapley2021a}. 

Formally, the Shapley value of a data point \(z_i \in D\) is given by:
\[
\phi_{z_i}(v) 
= \mathbb{E}_{D' \sim \mathcal{P}(D \setminus \{z_i\})} \bigl[v(D' \cup \{z_i\}) - v(D')\bigr],
\]
in which \(D'\) is a subset of \(D \setminus \{z_i\}\) sampled uniformly at random and \(v(D')\) is the value function evaluated on \(D'\). 
The Shapley value thus quantifies the \emph{expected} incremental contribution of each data point across all possible subsets. 
The uniqueness of the Shapley value can be derived from satisfying four foundational axioms: \emph{Efficiency} (\emph{i.e.}, the total value is distributed among all players), \emph{Symmetry} (\emph{i.e.}, identical contributions are rewarded equally), \emph{Dummy} (\emph{i.e.}, players that contribute nothing receive zero value) and \emph{Linearity} (\emph{i.e.}, the Shapley values from two games can be combined linearly). 
One application is \emph{data pricing} in marketplaces \cite{jiaEfficientDataValuation2019, peiSurveyDataPricing2022, tianPrivateDataValuation2022, xiaEquitableDataValuation2023}, in which Shapley-based frameworks assign higher prices to points that contribute across many subsets. 
Shapley values can also inform the \emph{data selection} task, by identifying the examples that are the most useful for training smaller high-quality sets. 
Their effectiveness depends on the structure of the utility function and faces the challenge of a high computational cost, with approximations such as G-Shapley \cite{ghorbani2019data, yoon2020data} proposed to improve scalability. 
However, for \(k\)-NN classifiers, an exact Shapley formulation exists \cite{jia2019efficient} with complexity \(O(N \log N)\), making valuation tractable in this setting.

\paragraph{Differential privacy.}
Differential Privacy (DP)~\cite{dwork2006differential} provides a mathematically rigorous framework for limiting how much information about an individual can be inferred from a released output.
In this work, we focus on the \emph{local setting}, in which each data point is randomized before it is disclosed to any external party \cite{cormode_privacy_2018, wang2019collecting}.
Formally, let \(z\) be an individual data point and let \(R\) be a randomized algorithm acting on single records.
We say that \(R\) satisfies \(\epsilon\)-local differential privacy if, for any pair of possible inputs \(z, z'\) and any measurable set of outputs \(O\),
\[
    \Pr[R(z) \in O] \le e^\epsilon \Pr[R(z') \in O].
\]
Classically, local DP is implemented in a \emph{decentralized} manner, by having each individual perturbing his own data before sending it to any curator.
Our setting is a bit different as the Data Union acts as a trusted curator and \emph{centrally} applies a LDP-mechanism at the record level by adding noise independently to each member's data point before releasing it to the Data Consumer.
Beyond protection, DP is also related to the incentives individuals face when deciding whether to share their data truthfully.
From a mechanism design perspective, the \textit{Revelation Principle}~\cite{myerson1983mechanism} states that outcomes achievable through indirect strategies can also be implemented by truthful revelation.
DP makes such mechanisms approximately truthful~\cite{mcsherry2007mechanism}, an \(\varepsilon\)-DP mechanism bounds each agent's influence on the outcome, so the \textit{gain from lying} is at most \(O(\varepsilon)\).

\section{Information Disclosure Game}
\label{sec:framework}

We model the interaction between two agents: the DU and the DC. 
More precisely, the DU manages pooled data with the mission of ensuring fair valuation while accounting for concerns such as privacy. 
Meanwhile, the DC seeks to optimize utility—often by acquiring data points at minimal cost. 
These interactions are formalized as a two-phase Stackelberg game, in which the DU acts as a leader by setting how to disclose information and at what cost, and the DC, as a follower, responds through strategic data acquisition.

\paragraph{Complete information disclosure game.}
A typical formulation is to model the interaction as a pricing game, in which the DU assigns a price $p_i$ to each data point $z_i$. 
The DC then solves a knapsack-like problem:
\begin{equation}
\min_{\mathbf{x} \in \{0,1\}^N} \;\; \sum_{i=1}^N p_i x_i 
\quad \text{s.t.} \quad U(\mathbf{x}) \ge U_{\text{target}},
\label{eq:pricing}
\end{equation}
in which $U(\mathbf{x})$ denotes the utility obtained from the purchased subset. 
The DU anticipates this behavior and sets $\mathbf{p}$ with the aim of maximizing the DC’s minimized total cost.

This approach provides a straightforward pricing mechanism but has significant limitations. 
Indeed, as the disclosure of a data point is binary—either withheld or complete—many contributors receive no compensation. 
Incentivizing broader acquisition requires lowering the price of high-value points, which penalizes those who contribute the most. 
Moreover, releasing data at fixed prices locks in value at a single point in time and weakens privacy guarantees, limiting the DU’s ability to balance inclusiveness and protection.

\paragraph{Partial information disclosure game.}
To address these challenges, we depart from per-point pricing and instead model an iterative disclosure process under DP. 
At each round $t$, the DC selects a subset $S_t$ of data points to query. 
Each query consumes a fixed privacy budget $\epsilon$ and returns a noisy version of the data. 
The DU sets $(\epsilon, T_{\max})$, the per-query budget and the per-point query cap, which together determine a maximum per-point spend $B_{\max} = T_{\max}\epsilon$. 
The DC’s cumulative spend is $B = \epsilon \sum_{t=1}^T |S_t|, \quad T \leq T_{\max}$.
The DC’s objective is to minimize its cumulative budget spent:
\begin{equation}
\min_{\{S_t\}} \; \epsilon \sum_{t=1}^{T} |S_t| 
\quad \text{s.t.} \quad U^{(T)} \ge U_{\text{target}}, \; T \le T_{\max}.
\end{equation}
Conversely, the DU’s objective is to maximize the minimum budget spent by the DC:
\begin{equation}
\max_{(\epsilon, T_{\max})} \; \min_{\{S_t\}} \epsilon \sum_{t=1}^T |S_t|,
\end{equation}
with the goal of spreading being spent more evenly across members while managing privacy–utility trade-offs.
Compared to the pricing game in Equation~\ref{eq:pricing}, partial disclosure replaces per-point ownership costs with iterative access charges. 
This formulation makes the hidden acquisition cost of valuation explicit in the DC’s budget and allows the DU to shape acquisition behaviors without penalizing the most valuable contributors.  

\section{Experiments on Review Helpfulness Prediction}
\label{sec:case_study}

Review helpfulness prediction provides a natural testbed for data valuation as platforms such as Yelp collect reviews not only for sentiment analysis but also to assess their quality.
Indeed, unhelpful reviews may ultimately be discarded, yet they are still necessary to identify and reward helpful ones.
This creates a setting in which individual contributions to predictive performance are both explicit and uneven. 
We use the Yelp dataset~\cite{asghar2016yelp} following the setup of~\citet{bilal2023effectiveness}, who report a $k$-NN accuracy of 59.6\%. 
Using pretrained text embeddings and $k$-NN, our best configuration achieves 66.0\% test accuracy and 65.2\% F1-score, bringing non-parametric performance closer to fine-tuned transformer models while preserving point-level interpretability needed for valuation.
As a sanity check, we replicated the acquisition experiment of~\citet{jia2019towards} and verified that Shapley-based selection outperforms random sampling. 
The corresponding curve, along with additional details on the experimental setup, is reported in Appendix~\ref{app:shapley_vs_random}.

To implement a DP iterative release, we add Laplacian noise independently to each feature of every data point (1024-dimensional vectors). 
For feature \(x_j\), we first project it into a fixed interval \([a_j,b_j]\), in which \(a_j\) and \(b_j\) are min/max statistics for feature \(j\). 
The release mechanism is then
\[
\tilde{x}_j \;=\; \min\{\max\{x_j,\,a_j\},\,b_j\} \;+\; \mathrm{Laplace}\!\left(\frac{\Delta_j}{\epsilon_j}\right),
\]
in which \(\Delta_j = b_j - a_j\) denotes the global sensitivity of feature \(j\) after bounding and \(\epsilon_j\) is the allocated budget for that feature. 
This design is similar in spirit to a per-feature local DP mechanism~\cite{cormode_privacy_2018, wang2019collecting}, in that each coordinate is privatized independently, but here it is applied in a \textit{centralized} setting by the DU. 

We assume the sensitivity interval \([a_j,b_j]\) is fixed in advance and does not vary significantly across releases, so the noise scale is determined by \(\Delta_j / \epsilon_j\). 
On the DC side, each noisy version of a point is averaged to form a denoised estimate. 
After observing \( t \) noisy versions of point \( i \), its center is computed incrementally as $\hat{x}_i^{(t)} = \frac{1}{t} \sum_{k=1}^t \tilde{x}_i^{(k)}$. 
The averaging strategy steadily improves fidelity as noisy samples accumulate, with error decreasing and correlation increasing non-linearly across iterations (see Appendix~\ref{app:fidelity} for details).  

In the remainder of our analysis, we do not attempt to solve the full game between the DU and DC. 
Instead, we simulate the DC’s behavior by implementing and comparing different data selection strategies under the noisy iterative release mechanism. 
We define the utility target as the validation accuracy (69.6\%). 
For all data selection strategies, we use a fixed privacy budget $\epsilon$ per point per query of 1 and assume a constant maximum budget per data point. 
The next two subsections explore these strategies, including one based on rankings (random and Shapley-based) and another using adaptive n-armed bandit algorithm.

\subsection{Data Selection Using Random and Shapley-based Strategies}

\begin{figure}[h!]
    \centering
    \begin{minipage}[t]{0.45\textwidth}
        \centering
        \includegraphics[width=\textwidth]{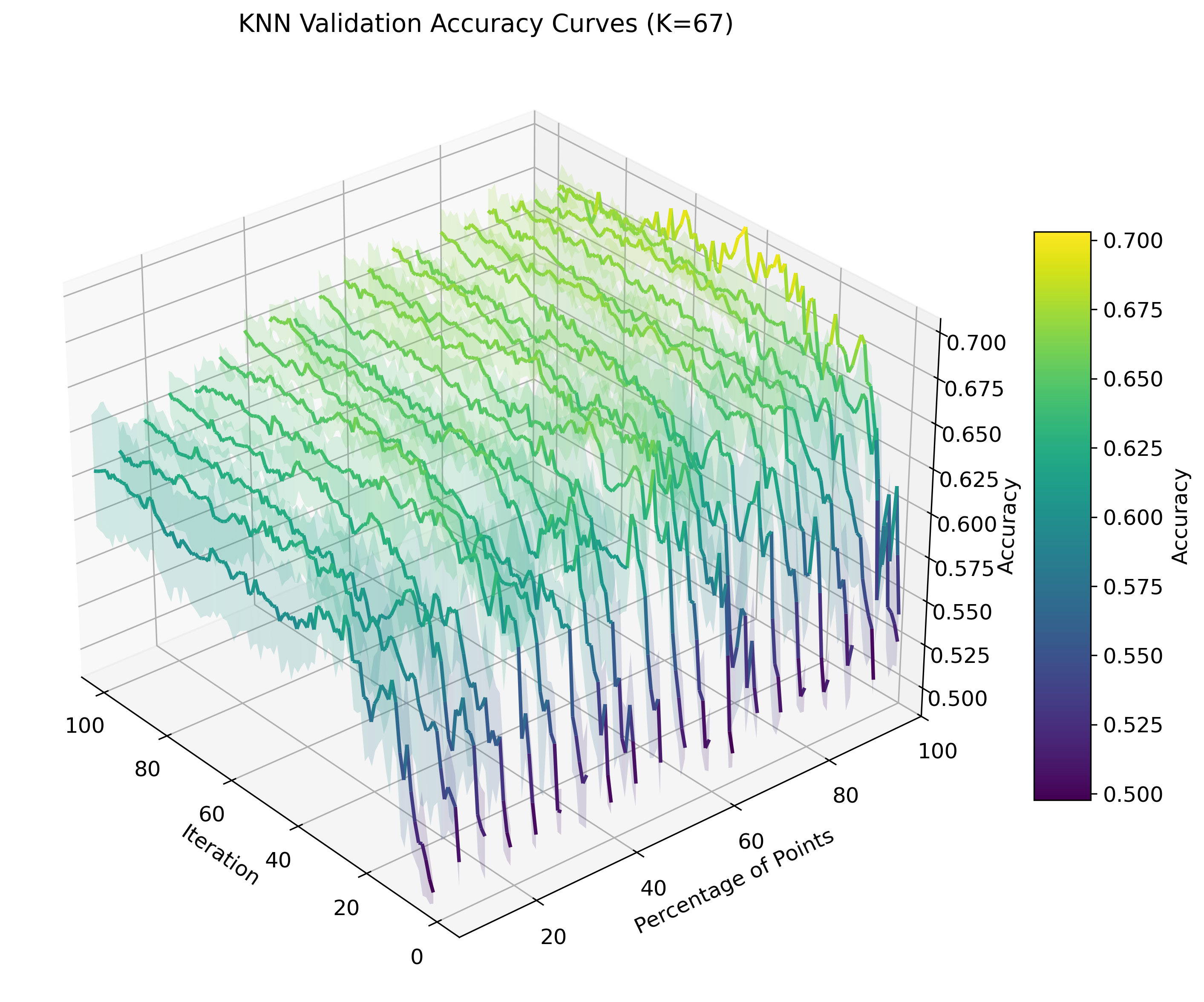}
        \caption{Validation accuracy for random data selection across varying parameters. 
        Unlike Shapley-based methods, random selection fails to consistently reach the utility target (69.6\%) within the budgeted iteration range.}
        \label{fig:random_curves_3d}
    \end{minipage}
    \hfill
        \begin{minipage}[t]{0.45\textwidth}
        \centering
        \includegraphics[width=\textwidth]{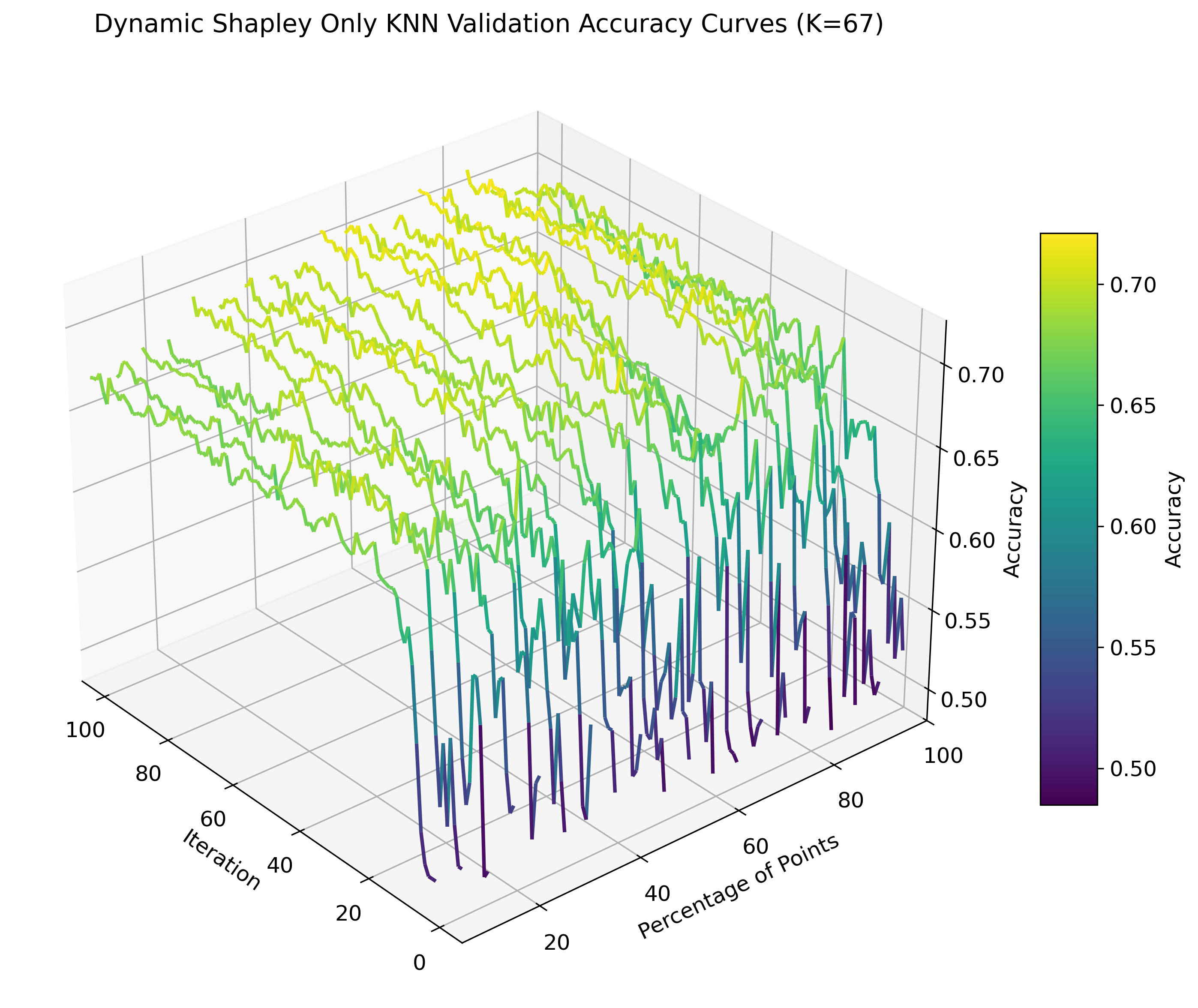}
        \caption{Validation accuracy using estimated data Shapley selection across dataset percentages and iterations. 
        The target accuracy is achieved with as little as 10\% of data.}
        \label{fig:shapley_curves_3d}
    \end{minipage}%

\end{figure}

\paragraph{Random data selection.} In this baseline, the DC commits to a random subset of data points for all iterations, with the results displayed in Figure~\ref{fig:random_curves_3d} being averaged over 10 random permutations. 
As shown in Figure~\ref{fig:random_curves_3d}, this approach fails to reach the target utility within 100 iterations unless nearly 100\% of the dataset is acquired, which confirms that a purely random acquisition strategy from the DU is not viable under budget constraints.

\paragraph{Data Shapley selection.} Afterwards, we assess whether Shapley valuation works over noisy data by the DC selecting all data points for a number of iterations (\emph{i.e.}, bootstrap iterations) and estimating DSVs based on the averaged center points. 
As illustrated in Figure~\ref{fig:shapley_curves_3d}, selecting the top-valued centers based on noisy Shapley estimates allows the DC to reach the target utility. 
Successful acquisition starts at 10\% of the dataset but performance varies depending on the percentage of selected data and the number of iterations. 
For instance, we observed that in the 60\% selection case, data Shapley selection achieves the utility target after roughly 25 bootstrap iterations. 
Testing other Shapley-based strategies (Appendix~\ref{app:shapley_vs_random}), we found no consistent advantage in committing earlier.
Although committing may stabilize performance for a fixed subset, our findings suggest that for Shapley-based strategies, the DC is better off continuing complete iterations to refine noisy point estimates.

\subsection{Data Selection using $n$-Armed Bandits}

As a final strategy, we model data selection as an $n$-armed bandit problem, in which each action $a \in \{1,\dots,n\}$ corresponds to selecting a data point \(z_a\). 
The DC follows an Upper Confidence Bound (UCB) policy under a fixed per-point privacy budget \(B_{\text{MAX}}\). 
Each query incurs a privacy cost $\epsilon$ and yields a noisy sample that updates the point’s estimated value, measured as incremental utility in a $k$-NN classifier. 
Unlike Shapley-based strategies, the DC may exploit the same point repeatedly until its budget is exhausted. 
This formulation captures the exploration–exploitation trade-off and makes the \emph{cost of valuation} explicit as the exploration cost required to identify valuable points under budget constraints. 
Formally, each point maintains its estimated value $Q_t(a)$, query count $N_t(a)$, current averaged center $\text{center}_a$ and remaining budget $B^{\text{re}}_a$. 
UCB scores are computed as
\[
\mathrm{UCB}_t(a) =
\begin{cases}
Q_t(a) + c \cdot \sqrt{\dfrac{1}{N_t(a) + \varepsilon}} \cdot \dfrac{B_a^{\text{re}}}{B_a^{\text{MAX}}}, & \text{if } B_a^{\text{re}} > 0 \\
-\infty, & \text{otherwise}.
\end{cases}
\]
The utility of a queried point is defined as the fraction of neighbors in $k$-NN that share its label. 
See Algorithm~\ref{alg:budget-ucb-knn} in appendices for implementation details.

\paragraph{Hyperparameter search.}  
Identifying an equilibrium can be seen as a hyperparameter search, since the interaction between the DU’s release policy and the DC’s acquisition strategy does not admit a closed-form solution. 
To realize this, we conducted a grid search over two key parameters: the \emph{maximum budget per point} (set by the DU) and the \emph{exploration coefficient} \(c\) (in UCB). 
These jointly determine budget usage and the diversity of selected points.  
Results show that the bandit reliably reaches the utility threshold except when the per-point budget is too small or exploration is disabled; below a budget of 20, success is rare and often due to \textit{lucky point selection}. 
Gini analysis of budget allocations confirms that exploration promotes more balanced spending across points and increases the likelihood of success. 
The correlation between Q values and Shapley values grows with the budget but remains moderate overall (Appendix~\ref{app:q-values-correlation}).
Finally, evaluations based on denoised centers yielded test accuracy comparable to or exceeding that obtained from the original data.
For instance, a configuration with a per-point budget of 50 reached 69.9\% surpassing the 66\% benchmark.

\begin{figure}[h!]
    \centering
    \begin{minipage}{0.45\textwidth}
        \centering
        \includegraphics[width=\linewidth]{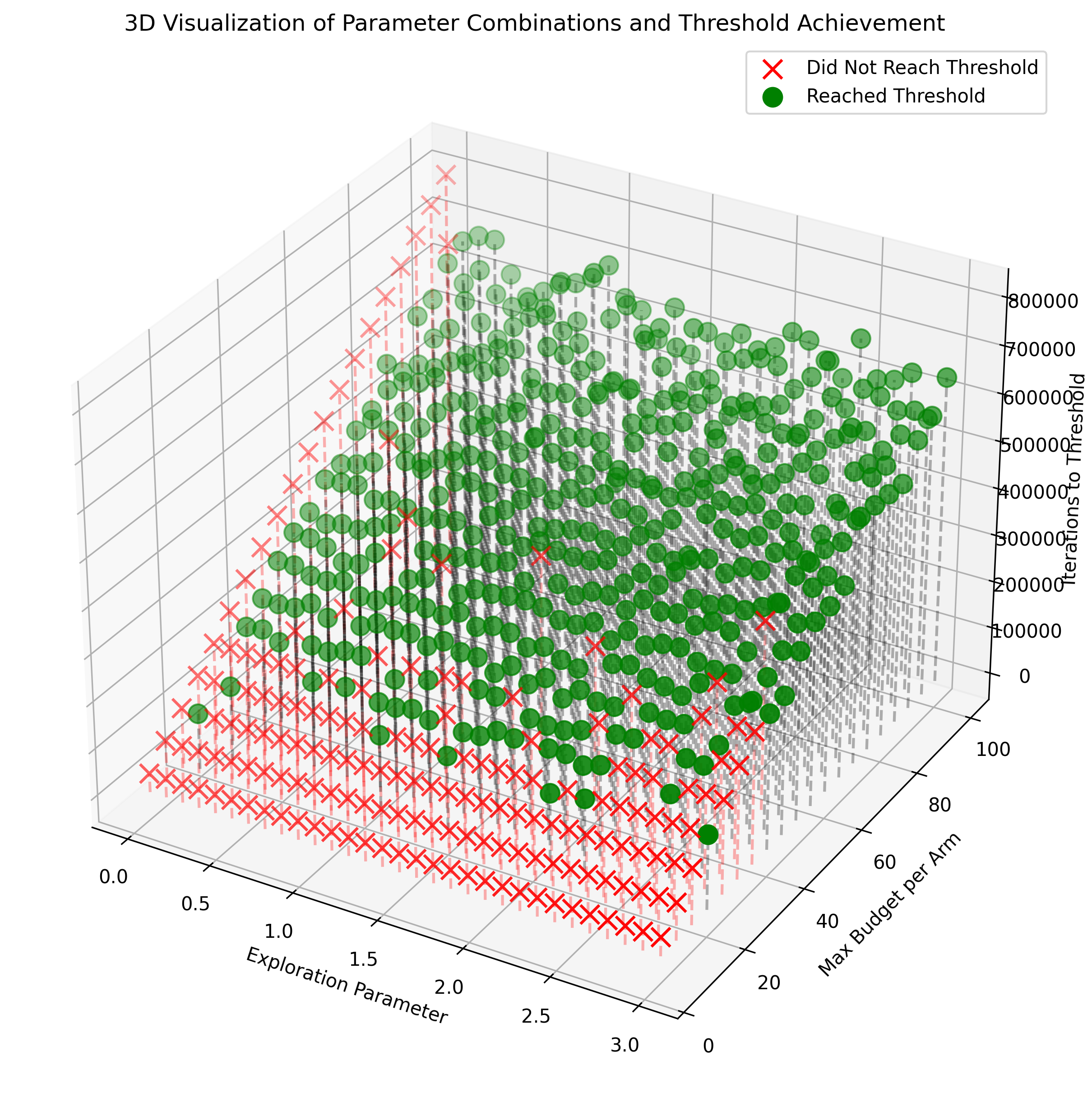}
        \caption{3D visualization of hyperparameter combinations. 
        Green dots represent successful runs in which the DC reached the utility threshold while red crosses represent failures. 
        Success becomes unlikely with budget-per-arm below 20 or when exploration is zero.}
        \label{fig:threshold_reached_3d}
    \end{minipage}\hfill
    \begin{minipage}{0.45\textwidth}
        \centering
        \includegraphics[width=\linewidth]{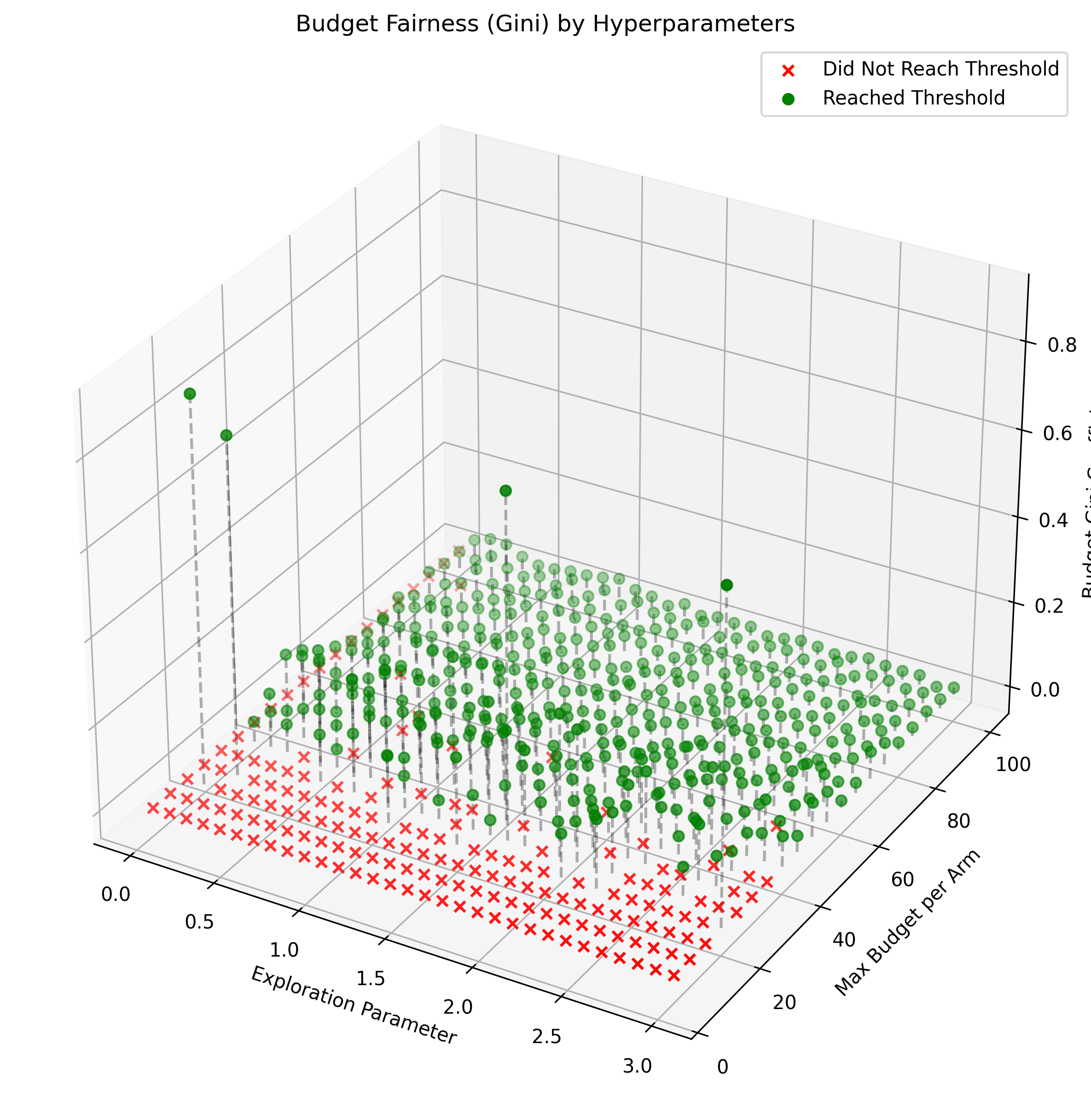}
        \caption{Gini coefficient of budget usage by hyperparameter setting. 
        High values indicate budget concentrated on few data points, typical in ``lucky'' early selections.}
        \label{fig:gini_budget_3d}
    \end{minipage}
\end{figure}

\section{Discussion \& Conclusion}
\label{sec:discussion_conclusion}

We proposed a game-theoretic framework for data valuation that integrates privacy and inclusiveness concerns through an information disclosure game between a Data Union (DU) and a Data Consumer (DC). 
Rather than assuming direct access to data, our approach models acquisition as an iterative process using a differentially-private mechanism to enforce a cost on data valuation. 
On Yelp review helpfulness prediction, Shapley-based strategies remained effective under noise but required a minimal budget—about 10 bootstrap iterations—to reliably identify high-value points. 
This demonstrates that Shapley valuation entails an inherent cost of exploration in our setting. 
Similarly, a multi-armed bandit strategy reached the utility target with comparable budgets, reinforcing that valuation itself imposes exploration costs. 
Gini analysis further showed that as budgets grow, the DC allocates resources more evenly across contributors, increasing inclusiveness.

While our experiments focused on $k$-NN for computation efficiency, an important direction for future work is extending the framework to differentiable models. 
Gradient-based methods such as \textit{G-Shapley}~\cite{ghorbani2019data} already show that Shapley values can be approximated from changes in gradients, suggesting ways to scale valuation beyond non-parametric models. 
In parallel, approaches like \textit{DP-SGD}~\cite{abadi2016deep} demonstrate that DP can be enforced by adding noise to gradients rather than directly to points. 
Bridging these lines of work would provide a better privacy-utility trade-off for both the DU and DC.
Nonetheless, our results suggest that in a DU setting, reaching target utility requires a minimum level of budget spread.
Thus, our framework implies that a \textit{minimum dividend} should be guaranteed to all members, regardless of their individual Shapley value. 
In the context of reviews, this means that every contributor would receive at least some share of value, even if their individual review is ultimately deemed unhelpful, aligning the incentives of all members with the collective outcome.
Otherwise, excluding some contributors increases the incentive to form data unions that adopt adversarial disclosure strategies (\emph{e.g.}, prioritizing low-value reviews or injecting excess noise), which may in turn hinder the data consumer’s ability to reach its utility target.

\bibliographystyle{abbrvnat}
\bibliography{biblio}
\newpage
\appendix

\section{Appendix: MAB Algorithm}
\label{app:map_algorithm}

\begin{algorithm}
\caption{Budget-Aware UCB for Data Selection}
\label{alg:budget-ucb-knn}
\begin{algorithmic}[1]
\Require 
\Statex $\mathcal{D} = \{x_1, \dots, x_n\}$: Dataset of $n$ data points (arms)
\Statex $\epsilon$: Differential privacy budget per query
\Statex $T_{\max}$: Maximum number of iterations
\Statex $U_{\text{target}}$: Target utility (average positive vote ratio)
\Statex $c$: Exploration coefficient
\Statex $\alpha$: Learning rate
\Statex $\varepsilon$: Small constant for numerical stability
\Statex $B^{\text{MAX}}(a)$: Maximum privacy budget per point
\Statex \textsc{NoisyRelease}($x_a$, $\epsilon$): Returns a noisy version of $x_a$
\Statex \textsc{knnUtility}($\mathcal{Z}$; $\{\text{center}_a\}, k$): Average positive votes ratio over validation set $\mathcal{Z}$ with current centers

\State \textbf{Initialize:}
\For{$a = 1$ \textbf{to} $n$}
    \State $Q(a) \gets 0$, $N(a) \gets 0$, $B^{\text{re}}(a) \gets B^{\text{MAX}}(a)$
    \State $\text{center}_a \gets 0$
\EndFor
\State $U \gets \textsc{knnUtility}(\mathcal{Z}; \{\text{center}_a\}, k)$
\State $t \gets 1$

\While{$t \leq T_{\max}$ \textbf{and} $U < U_{\text{target}}$}
    \For{$a = 1$ \textbf{to} $n$}
        \If{$B^{\text{re}}(a) > 0$}
            \State $\mathrm{UCB}(a) \gets Q(a) + c \cdot \sqrt{\frac{1}{N(a) + \varepsilon}} \cdot \frac{B^{\text{re}}(a)}{B^{\text{MAX}}(a)}$
        \Else
            \State $\mathrm{UCB}(a) \gets -\infty$
        \EndIf
    \EndFor

    \State $A_t \gets \arg\max_{a} \mathrm{UCB}(a)$
    \State $\tilde{x}_{A_t} \gets \textsc{NoisyRelease}(x_{A_t}, \epsilon)$
    \State $\text{center}_{A_t} \gets \frac{\text{center}_{A_t} \cdot N(A_t) + \tilde{x}_{A_t}}{N(A_t) + 1}$

    \State $U_{\text{new}} \gets \textsc{knnUtility}(\mathcal{Z}; \{\text{center}_a\}, k)$
    \State $R_t \gets U_{\text{new}} - U$
    \State $U \gets U_{\text{new}}$

    \State $Q(A_t) \gets Q(A_t) + \alpha \cdot (R_t - Q(A_t))$
    \State $N(A_t) \gets N(A_t) + 1$
    \State $B^{\text{re}}(A_t) \gets B^{\text{re}}(A_t) - \epsilon$
    \State $t \gets t + 1$
\EndWhile

\State \Return $\{\text{center}_a\}, U$
\end{algorithmic}
\end{algorithm}

\section{Appendix: The effect of Laplacian noise}

We have allocated a privacy budget of \( \epsilon = 1 \) per feature, resulting in a total budget of 1024 per point. 
To illustrate how fidelity evolves under iterative disclosure, we report two complementary measures. 
The first tracks how close denoised representations remain to the original data points, while the second measures the correlation between original Shapley values and those computed on denoised representations. 
Together, they provide evidence that averaging across noisy samples progressively restores signal.

\label{app:fidelity}
\begin{figure}[h]
    \centering
    \includegraphics[width=0.6\textwidth]{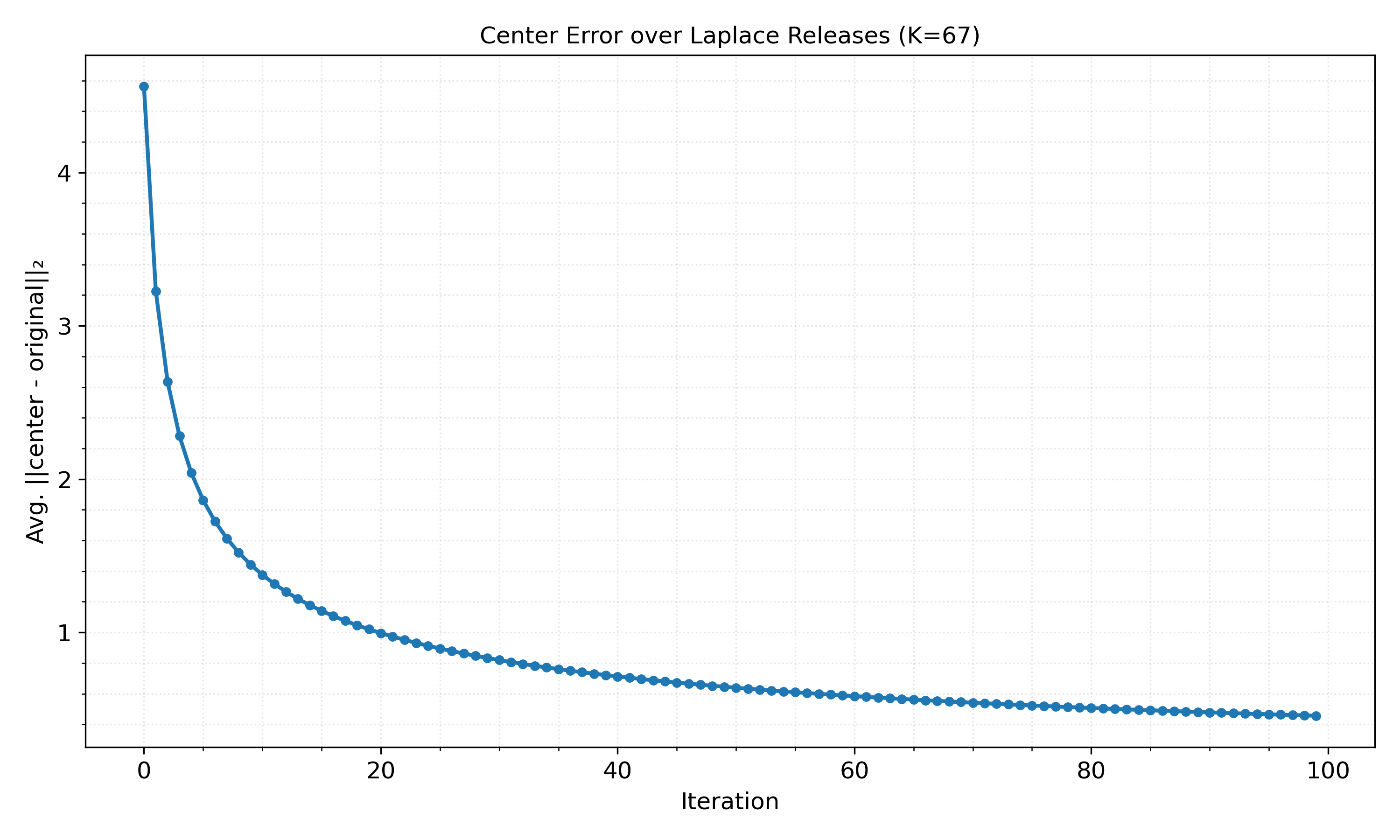}
    \caption{Average \( \ell_2 \) distance between original points and their denoised centers as a function of iterations. 
    Error drops sharply early on with diminishing returns over time.}
    \label{fig:center_distance}
\end{figure}

\begin{figure}[h]
    \centering
    \includegraphics[width=0.6\textwidth]{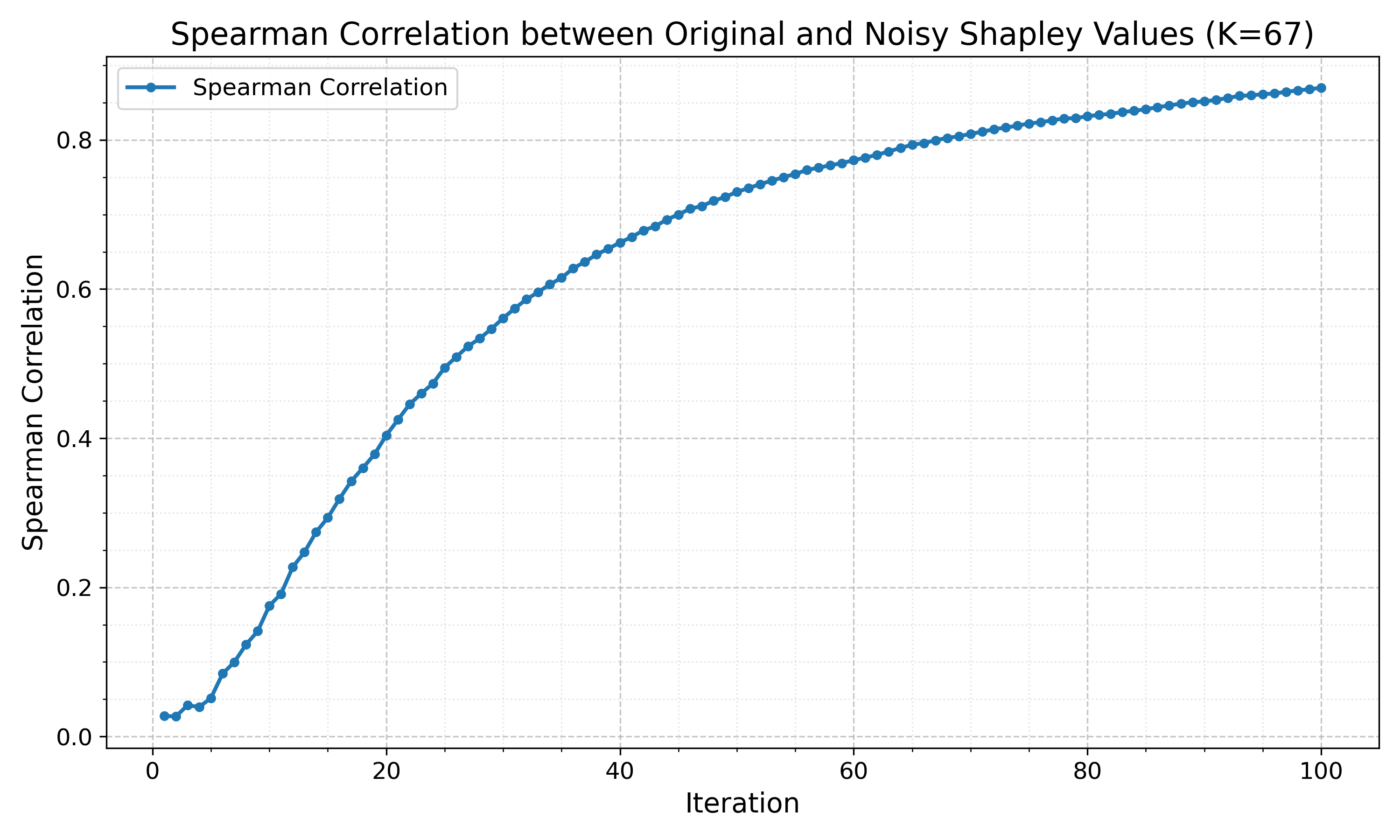}
    \caption{Spearman correlation between original Shapley values and those computed on denoised centers over iterations.}
    \label{fig:spearman_corr}
\end{figure}

\section{Appendix: Experimental setup}
\label{app:experimental_setup}

\paragraph{Dataset and embeddings.} 
We used the Yelp dataset~\cite{asghar2016yelp}, which contains text reviews with helpfulness votes. 
Reviews were encoded with two sentence embedding models: \texttt{sentence-transformers/all-mpnet-base-v2} and \texttt{intfloat/multilingual-e5-large-instruct}. Both were chosen for their strong performance on semantic similarity tasks and because they were not trained on Yelp, avoiding leakage. 
A $k$-Nearest Neighbors classifier was trained with the number of neighbors $k$ tuned on a validation set. 
The dataset was split into 8000 training, 1000 for validation and 1000 test examples.

\paragraph{Computing resources.}  
All experiments were run locally on a MacBook Pro M4 with 24GB of unified memory. 
Each acquisition experiment completed within a few hours, with bandit simulations being the most computationally intensive.

\begin{figure}[h!]
    \centering
    \includegraphics[width=0.5\textwidth]{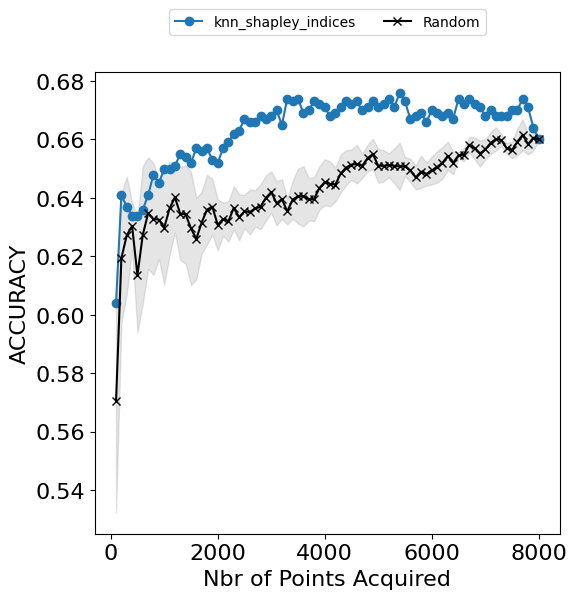}
    \caption{Test accuracy as a function of acquired data points, comparing kNN Shapley-based selection with random selection.}
    \label{fig:shapley_vs_random_app}
\end{figure}

\section{Appendix: Data Shapley Selection strategies}
\label{app:shapley_vs_random}

\begin{figure}[h!]
    \centering
    \includegraphics[width=0.5\textwidth]{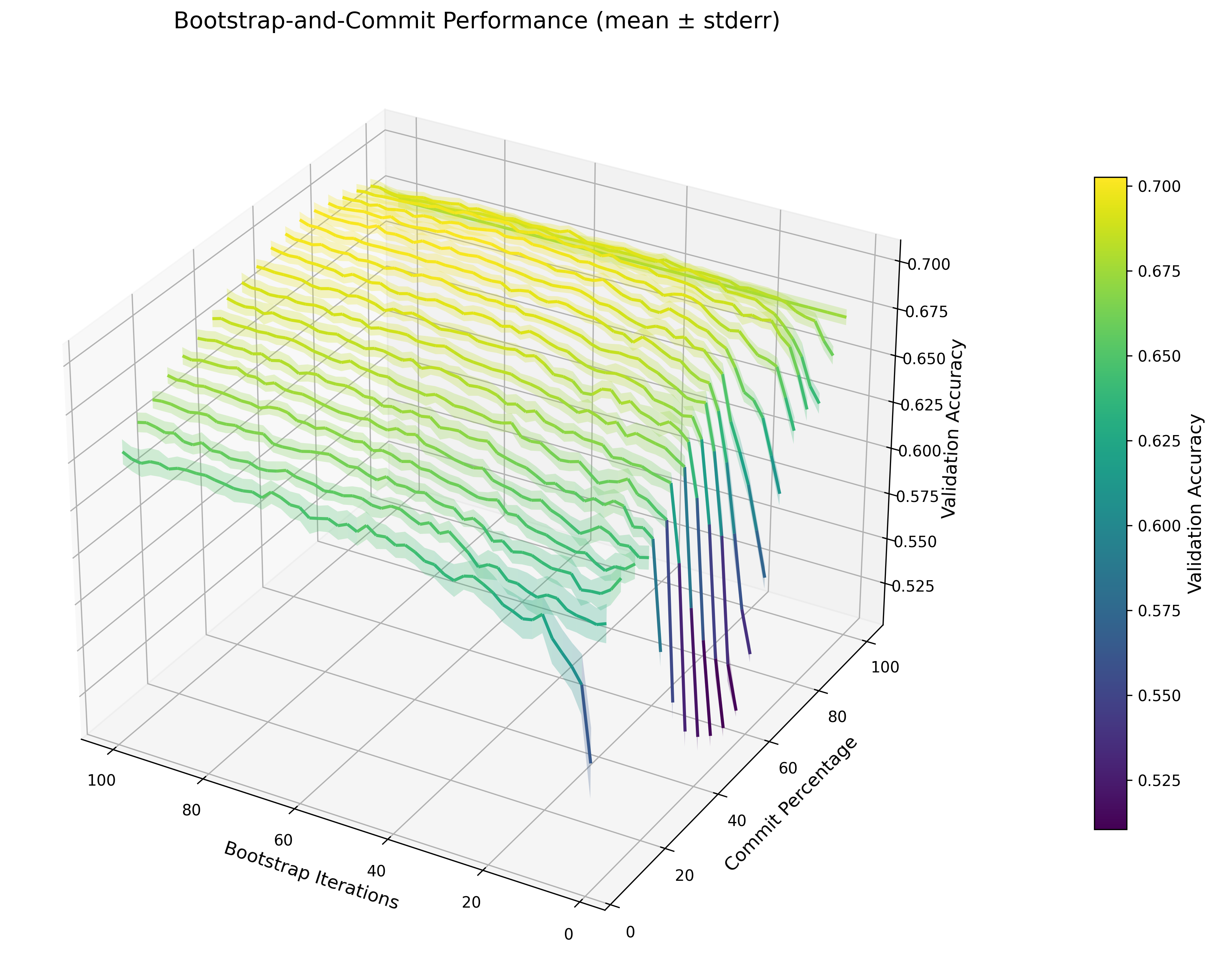}
    \caption{Performance of data Shapley+commitment selection strategy for different combinations of bootstrap and commit parameters. 
    No clear benefit is observed compared to no commitment.}
    \label{fig:bootstrap_commit_3d}
\end{figure}

\begin{figure}[h!]
    \centering
    \includegraphics[width=0.5\textwidth]{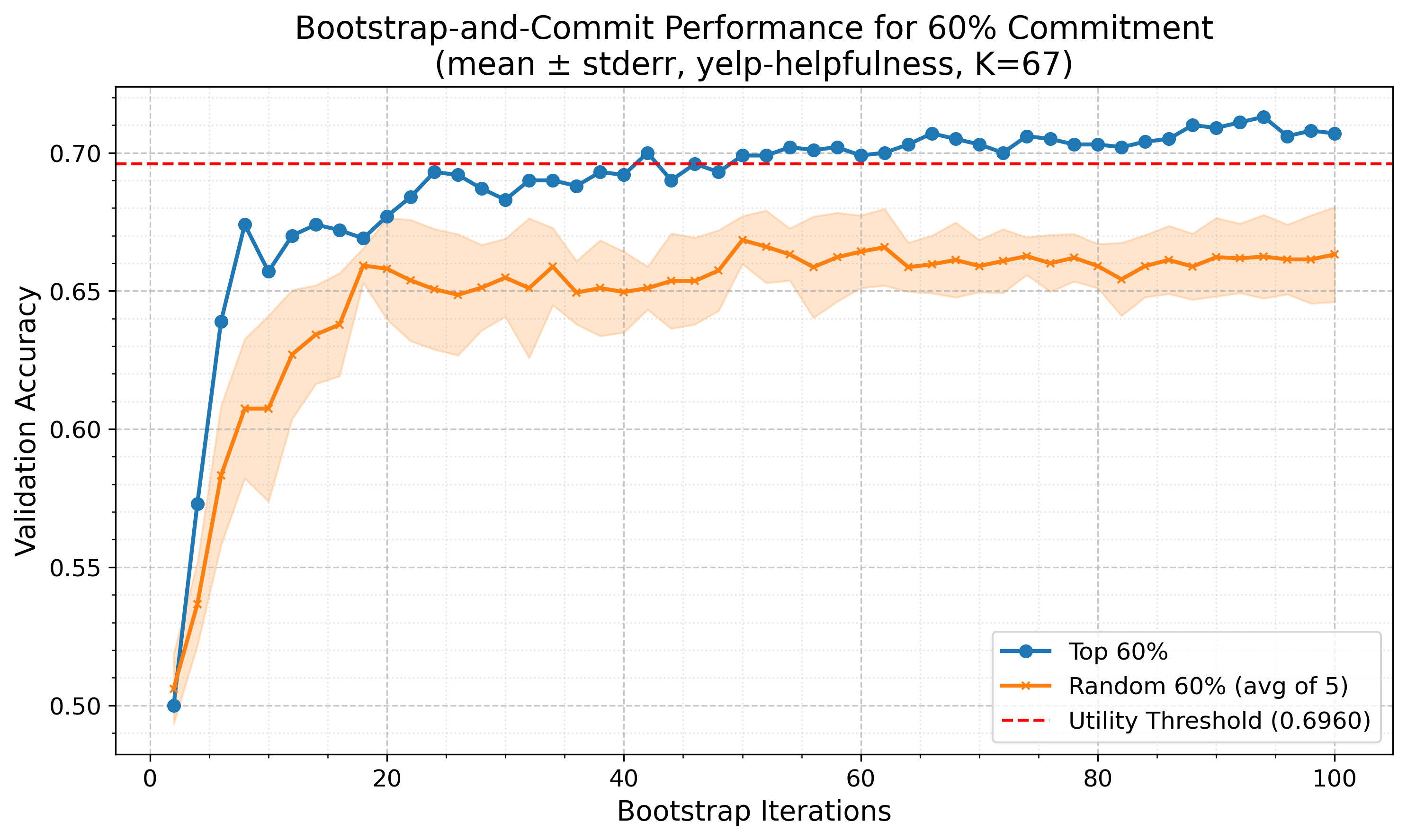}
    \caption{Shapley+commitment strategy on 60\% of the data. 
    While utility is eventually reached, performance does not improve over dynamic Shapley.}
    \label{fig:bootstrap_commit_60pct}
\end{figure}

Figure~\ref{fig:bootstrap_commit_3d} illustrates an alternative acquisition strategy in which the DC first estimates Shapley values through a fixed number of bootstrap iterations before committing to acquire the top-ranked points. 
This procedure is compared against the exact Shapley values provided by an oracle, which serves as a benchmark but is not available in practice. 
The results highlight that while estimating Shapley values incurs an additional cost, this cost diminishes as more budget is consumed. 
However, committing after a fixed iteration number does not reduce the total number of iterations required to reach the target utility, as shown in Figure~\ref{fig:bootstrap_commit_60pct}, indicating that early commitment provides no fundamental shortcut in convergence.

\begin{figure}[h!]
    \centering
    \includegraphics[width=0.5\textwidth]{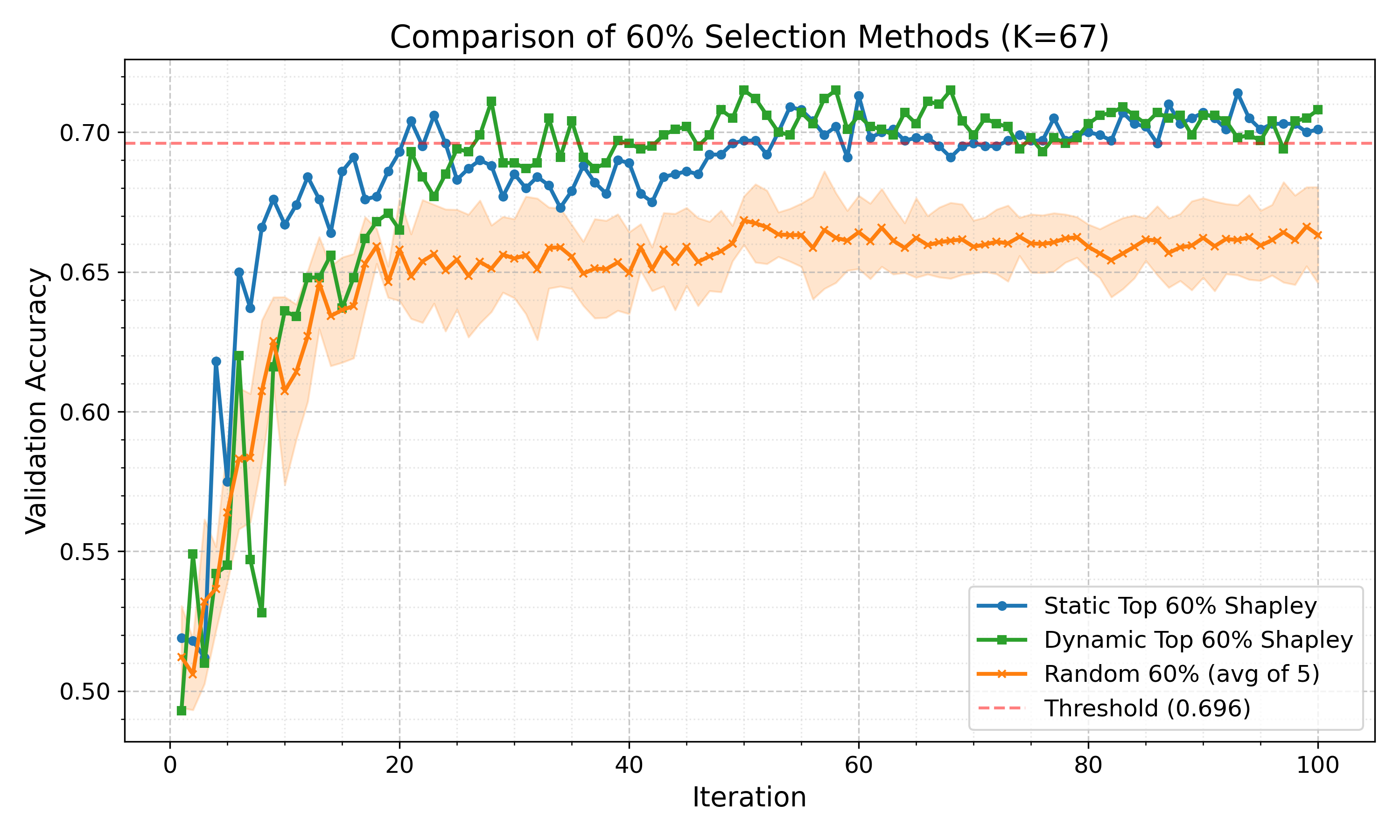}
    \caption{Validation accuracy over iterations for 60\% data selection using random, esimated (noisy) data Shapley strategies compared to exact Shapley values given by an oracle.
    Estimated data Shapley selection reaches the utility target in ~25 iterations, significantly outperforming random selection.}
    \label{fig:60pct_methods}
\end{figure}

\section{Appendix: Correlation between Q-values and Shapley values}
\label{app:q-values-correlation}
\begin{figure}[H]
    \centering
    \includegraphics[width=0.5\textwidth]{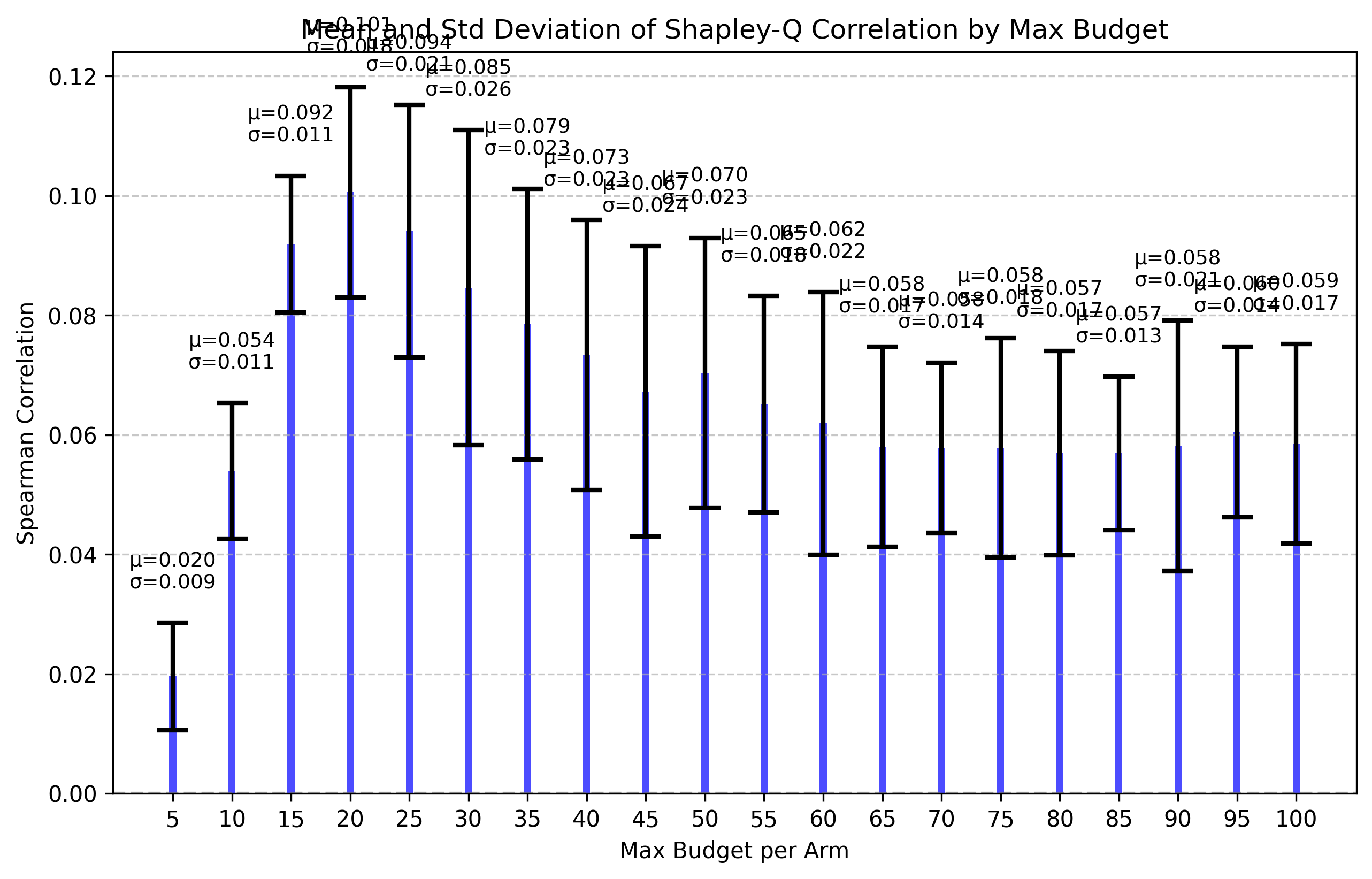}
    \caption{Mean and standard deviation of Spearman correlation between learned Q-values and Shapley values, grouped by maximum budget. 
    Correlation rises with budget but remains modest indicating partially aligned but distinct prioritization.}
    \label{fig:shapley_q_corr}
\end{figure}

We examined the relationship between learned Q-values and Shapley values. 
In our framework, Q-values are estimates maintained by the multi-armed bandit policy, representing the expected incremental utility of querying a particular data point under the differential privacy budget. 
Figure~\ref{fig:shapley_q_corr} shows that the Spearman correlation between these metrics increases with budget but remains moderate overall. 
This is expected and desired as the objective is for Q-values to be influenced by data utility but not perfectly mimic Shapley values, thereby offering a different valuation with more inclusiveness.

\end{document}